%% file: paper.tex
\pdfoutput=1
\documentclass[letterpaper,twocolumn,10pt]{article}

\input{texSetup}
\input{glossary}
\usepackage[normalem]{ulem}

\begin{document}

\title{Kernel-as-a-Service: A Serverless Interface to GPUs}

\author{Nathan Pemberton}
\author{Anton Zabreyko}
\author{Zhoujie Ding}
\author{Randy Katz}
\author{Joseph Gonzalez}
\affil{UC Berkeley}

\date{}

\begin{titlingpage}
    \maketitle
    \begin{abstract}
    \import{mainSrc/}{abstract}

    \end{abstract}
\end{titlingpage}

\glsresetall

\levelstay{Introduction}
\label{sec:intro}
\subimportlevel{mainSrc/}{intro}{1}

\levelstay{Cloud GPUs Today}
\label{sec:currentAproaches}
\subimportlevel{mainSrc/}{currentAproaches}{1}

\levelstay{A New Approach: Kernel-as-a-Service}
\label{sec:kaas}
\subimportlevel{mainSrc/}{kaas}{1}

\levelstay{Implementation}
\label{sec:implementation}
\subimportlevel{mainSrc/}{implementation}{1}

\levelstay{Evaluation}
\label{sec:evaluation}
\subimportlevel{mainSrc/}{evaluation}{1}

\levelstay{Related Work}
\label{sec:related}
\subimportlevel{mainSrc/}{relatedWork}{1}

\levelstay{Resource-Specialized Function Types}
\label{sec:future}
\subimportlevel{mainSrc/}{future}{1}

\FloatBarrier
\levelstay{Conclusion}
\label{sec:conclusion}
\subimportlevel{mainSrc/}{conclusion}{1}


\FloatBarrier
\newpage
\printbibliography

\end{document}

%% file: texSetup.tex
\usepackage{import}
\usepackage{modular}
\usepackage[binary-units,per-mode=symbol]{siunitx}
\usepackage{placeins}
\usepackage{stfloats}
\usepackage{caption}
\usepackage{subcaption}
\usepackage[export]{adjustbox}
\usepackage{lipsum}
\usepackage[maxbibnames=100,style=numeric]{biblatex}
\usepackage{tabularx}
\usepackage{makecell}
\usepackage{authblk}
\usepackage{titling}
\usepackage{algorithm}
\usepackage{algpseudocode}
\usepackage{hyperref}
\usepackage{xcolor}
\usepackage[acronym, nopostdot, noredefwarn]{glossaries}
\glsdisablehyper

\usepackage[left=0.75in, right=0.75in, top=1in, bottom=1.5in]{geometry}
\hypersetup{
    colorlinks,
    linkcolor={red!50!black},
    citecolor={blue!50!black},
    urlcolor={blue!80!black}
}

\newcommand{\az}[1]{}

%% file: glossary.tex
\makeglossaries

\newcommand*{\newdualentry}[5][]{%
  \newglossaryentry{main-#2}{name={#4},%
  text={#3\glsadd{#2}},%
  description={#5},%
  #1  
  }%
  \newacronym{#2}{#3\glsadd{main-#2}}{#4}%
}

\newdualentry{wsc}
  {WSC}
  {warehouse-scale computer}
  {Generic term referring to tightly-integrated clusters of
    machines deployed in the datacenter.}

\newglossaryentry{sip}
{
  name={SiP},
  description={system in package},
  first={\glsentrydesc{sip} (\glsentrytext{sip})},
  plural={SiPs},
  descriptionplural={systems in package},
  firstplural={\glsentrydescplural{sip} (\glsentryplural{sip})}
} 

\newglossaryentry{soc}
{
  name={SoC},
  description={system on chip},
  first={\glsentrydesc{soc} (\glsentrytext{soc})},
  plural={SoCs},
  descriptionplural={systems on chip},
  firstplural={\glsentrydescplural{soc} (\glsentryplural{soc})}
} 

\newglossaryentry{faas}
{
    name={FaaS},
    description={Function-as-a-Service: A programming style focused on explicit state and ephemeral functions that operate on that state.},
    first={Function-as-a-Service (\glsentrytext{faas})}
}

\newglossaryentry{chipyard}
{
    name={Chipyard},
    description={A system-on-chip development framework developed at UC Berkeley.}
}

\newglossaryentry{paging}
{
  name={paging},
  description={The process of storing logical pages on an external storage device in order to free physical memory. Also called "swapping".}
}

\newdualentry{pfa}
  {PFA}
  {page fault accelerator}
  {The proposed hardware-accelerator that handles page-faults for
  remote pages automatically.}

\newglossaryentry{disag}
{
  name={dissaggregation},
  description={The WSC design that moves compute resources (such as memory,
disk, and CPUs) into dedicated resource-blades that are connected through a
high-performance network.}
}

\newdualentry{numa}
  {NUMA}
  {non-uniform memory access}
  {A system where memory is cache-coherently available to multiple CPUS, but
   with varying access latencies and bandwidths (a type of multi-socket
   machine).}

\newdualentry{rdma}
  {RDMA}
  {remote direct memory access}
  {A system where memory is directly addressable between multiple
   nodes through a network interface. RDMA systems are not typically
   cache-coherent.}

\newglossaryentry{page}
{
  name={page},
  description={A fixed-size logical group of data (typically
\SI{4}{\kibi\byte}). Sometimes called ``virtual page''.}
}

\newglossaryentry{page frame}
{
  name={page frame},
  description={A fixed-size region of physical memory used to store pages.}
}

\newglossaryentry{frame}
{
  name={frame},
  description={Synonym for page frame},
  see=[Glossary:]{page frame}
}

\newglossaryentry{pgtbl}
{
  name={page table},
  description={A hardware-visible tree in main memory that contains
translations from virtual to physical addresses.}
}

\newdualentry{pte}
  {PTE}
  {page table entry}
  {A single entry of the page-table. Each PTE refers to a single
  virtual page.}

\newdualentry{tlb}
  {TLB}
  {translation look-aside buffer}
  {A cache of virtual to physical address translations.}

\newdualentry{ptw}
  {PTW}
  {page table walker}
  {A hardware device to automatically walk the page-table and
  locate PTEs for a particular virtual address.}

\newglossaryentry{bookkeeping}
{
  name={bookkeeping},
  description={The internal OS-specific tasks related to bringing in a new
page. This typically includes updating meta-data and other page-tracking
activities.}
}

\newglossaryentry{freeq}
{
  name={FreeQ},
  description={Queue of free frames to be used by the PFA to service
page-faults.}
}

\newglossaryentry{newq}
{
  name={NewQ},
  description={Queue of new-page descriptors populated by the PFA on every page
fault and drained by the OS for bookkeeping.}
}

\newglossaryentry{evictq}
{
  name={EvictQ},
  description={Queue of pages to be evicted by the PFA. Populated by the OS
when it needs to free local physical memory.}
}

\newglossaryentry{pgid}
{
  name={pageID},
  description={A unique identifier for a page in remote memory. Acts as a
remote-memory address.}
}

\newglossaryentry{memory blade}
{
  name={memory blade},
  description={A dedicated memory node in a disaggregated system. The memory blade
exists solely to server memory requests. A memory blade may be custom-designed
for this purpose, or may simply expose an RDMA interface.}
}

\newdualentry{mtu}
  {MTU}
  {maximum transfer unit}
  {The largest contiguous packet that a network is capable of
  transmitting.}

\newdualentry{tmem}
  {TMem}
  {transcendent memory}
  {A layer in the Linux paging subsystem that stores pages in
  specialized memory that may not be disk-backed.}

\newglossaryentry{cgroup}
{
  name={cgroup},
  first={control group (cgroup)},
  description={The per-task (or group of tasks) resource management system in
the Linux kernel.}
}

\newglossaryentry{swap}
{
  name={swap},
  description={Historically used to refer to the process of moving an entire
process's memory image to disk, Linux uses ``swapping'' to refer to all
paging.}
  see=[Glossary:]{paging}
}

\newglossaryentry{anonpg}
{
  name={anonymous page},
  description={A page that does not contain disk-backed information. This is
primarily ``heap'' memory (e.g. memory allocated through malloc())}
}

\newdualentry{vma}
  {VMA}
  {virtual memory area}
  {Contiguous region of virtual memory used by Linux to simplify memory
   management.}

\newglossaryentry{swpent}
{
  name={swap entry},
  description={A Linux-specific value stored in evicted PTEs that contains information on where to locate an evicted page.}
}

\newglossaryentry{kpfad}
{
  name={kpfad},
  description={A background daemon that opportunistically performs bookkeeping
  and maintenance for the PFA.}
}

\newglossaryentry{kswapd}
{
  name={\texttt{kswapd}},
  description={A background kernel thread that opportunistically performs
    bookkeeping.}
}

\newglossaryentry{task}
{
  name={task},
  description={Linux kernel internal abstraction of a process.}
}

\newdualentry{lru}
  {LRU}
  {least recently used}
  {An algorithm that attempt to pick pages that have not been used recently.}

\newdualentry{nvm}
  {NVM}
  {non-volatile memory}
  {Storage devices with near-DRAM performance, and byte addressablity, that do
  not lose their data when powered off.}

\newdualentry{tco}
  {TCO}
  {total cost of ownership}
  {A metric that includes not only up-front costs of a system, but the total
  cost to own and operating that system for its effective lifespan.}

\newdualentry{pgas}
    {PGAS}
    {partitioned global address space}
    {A language-based technique that partitions the program's address space between local and remote objects.}
    
\newdualentry{dsm}
    {DSM}
    {distributed shared memory}
    {A broad class of techniques to present the illusion of a single address space across a distributed system}
    
\newdualentry{kaas}
    {KaaS}
    {Kernel-as-a-Service}
    {My proposed GPU-specific extension to the FaaS paradigm.}
    
\newglossaryentry{actor}
{
    name={actor},
    description={A persistent resource allocation associated with high-level user code (as opposed to an explicit container or virtual machine)}
}

\newdualentry{gtask}
    {eTask}
    {Exclusive Task}
    {A FaaS+GPU approach where serverless functions run with exclusive access to a GPU.}
    
\newdualentry{ktask}
    {kTask}
    {kernel task}
    {The GPU-specialized serverless function type proposed by KaaS}
    
\newdualentry{ctask}
    {cTask}
    {CPU task}
    {Traditional CPU-oriented serverless functions}
    
\newglossaryentry{func}
{
    name={function},
    description={A logical transformation over data that can be executed by a serverless framework. Functions can be implemented in many ways.}
}

\newglossaryentry{tail}
{
    name={tail latency},
    description={The worst case latency of some task. These are typically measured as percentiles (i.e., 99th percentile latency).}
}

\newdualentry{ddc}
    {DDC}
    {disaggregated datacenter}
    {A warehouse-scale computer design that physically deploys individual resources as globally accessible network-attached components.}
    
\newdualentry{pcsi}
    {PCSI}
    {portable cloud system interface}
    {A hypothetical unified system interface to warehouse-scale computers based on serverless computing.}
    
\newdualentry{api}
    {API}
    {application programming interface}
    {A standard set of user-controllable features that a system supports}
    
\newdualentry{slo}
    {SLO}
    {service-level objective}
    {A target latency for an operation below which the user experiences little additional benefit.}
    
\newdualentry{nic}
    {NIC}
    {network interface card}
    {The hardware device providing network connectivity. This term is used even if the network device is not implemented as a discrete external card.}

\newglossaryentry{cuda}
{
    name={CUDA},
    description={NVidia's GPU programming language},
}

%% file: mainSrc/abstract.tex
Serverless computing has made it easier than ever to deploy applications over scalable cloud resources, all the while driving higher utilization for cloud providers. While this technique has worked well for easily divisible resources like CPU and local DRAM, it has struggled to incorporate more expensive and monolithic resources like GPUs or other application accelerators. We cannot simply slap a GPU on a FaaS platform and expect to keep all the benefits serverless promises. We need a more tailored approach if we want to best utilize these critical resources. 

In this paper we present \gls{kaas}, a serverless interface to GPUs. In \gls{kaas}, GPUs are first-class citizens that are invoked just like any other serverless function. Rather than mixing host and GPU code as is typically done, \gls{kaas} runs graphs of GPU-only code while host code is run on traditional functions.  The \gls{kaas} system is responsible for managing GPU memory and schedules user kernels across the entire pool of available GPUs rather than relying on static allocations. This approach allows us to more effectively share expensive GPU resources, especially in multitenant environments like the cloud. We add support for KaaS to the Ray distributed computing framework and evaluate it with workloads including a TVM-based deep learning compiler and a BLAS library. Our results show that \gls{kaas} is able to drive up to 50x higher throughput and 16x lower latency when GPU resources are contended.